\def\selectedoptions{}
\ifx\empty\selectedoptions
  \def\selectedoptions{final}
\fi

\documentclass[
   \selectedoptions
  ]
  {aipproc}

\def\selectedlayoutstyle{6x9} 
\layoutstyle\selectedlayoutstyle

\SetInternalRegister\hbadness{8000} 

\newcommand\doingARLO[2][]{%
  \ifx\mmref\undefined #1\else #2\fi
}

\begin{document}

\title
[]
{CKM Matrix: Status and New Developments}

\author{A.~H\"ocker}{
  address={Laboratoire de l'Acc\'el\'erateur Lin\'eaire, IN2P3-CNRS et Universit\'e de Paris-Sud, BP 34, F-91405 Orsay Cedex, France}
}
\author{\underline{H.~Lacker}}{
  address={Laboratoire de l'Acc\'el\'erateur Lin\'eaire, IN2P3-CNRS et Universit\'e de Paris-Sud, BP 34, F-91405 Orsay Cedex, France}
}
\author{S.~Laplace}{
  address={Laboratoire de l'Acc\'el\'erateur Lin\'eaire, IN2P3-CNRS et Universit\'e de Paris-Sud, BP 34, F-91405 Orsay Cedex, France}
}
\author{F.~Le Diberder}{
  address={Laboratoire de l'Acc\'el\'erateur Lin\'eaire, IN2P3-CNRS et Universit\'e de Paris-Sud, BP 34, F-91405 Orsay Cedex, France}
}
\copyrightyear  {2001}

\begin{abstract}
An analysis of the CKM matrix parameters within the {\it R}fit
approach is presented using updated input values with special 
emphasis on the recent $\sin{2\beta}$ measurements from BABAR 
and Belle.  
The QCD Factorisation Approach describing $B \rightarrow \pi\pi,K\pi$ 
decays has been implemented in the software package CKMfitter. 
Fits using branching ratios and CP asymmetries are discussed.
\end{abstract}
\date{\today}
\maketitle
\section{Statistical Framework and Inputs}\label{sec:statistics}
In the Standard Model (SM) with three families, CP violation is 
generated by a single phase in the CKM matrix\,\cite{KM}. This 
picture can be probed quantitatively by means of a global fit to 
all quantities sensitive to CKM elements in the SM. The analysis 
presented here is performed within the {\it R}fit statistical 
approach\,\cite{HLLL}, which is implemented in the software package 
CKMfitter\,\cite{webpage}.
\\ 
The quantity $\chi^2 = -2 \ln{{\cal L}(y_{\rm mod})}$ is minimized 
in the fit, 
where the likelihood function is defined by
     ${\cal L}(y_{\rm mod}) 
    = {\cal L}_{\rm exp}(x_{\rm exp}-x_{\rm theo}(y_{\rm mod})) 
\cdot {\cal L}_{\rm theo}(y_{\rm QCD})$. 
The experimental part, ${\cal L}_{\rm exp}$, depends on measurements, 
$x_{\rm exp}$, and theoretical predictions, $x_{\rm theo}$, which are 
functions of model parameters, $y_{\rm mod}$. The theoretical part, 
${\cal L}_{\rm theo}$, describes the knowledge on the QCD parameters, 
$y_{\rm QCD} \in\{ y_{\rm mod} \}$, where the theoretical uncertainties 
are considered as allowed ranges. 
The agreement between data and the SM is gauged by the global minimum 
$\chi^2_{{\rm min};y_{\rm mod}}$, determined by varying all model 
parameters $y_{\rm mod}$. For $\chi^2_{{\rm min};y_{\rm mod}}$, a 
confidence level (CL), expressing the goodness-of-fit, is computed by 
means of a Monte Carlo simulation.
\begin{table}
\begin{tabular}{cccc}
 & &  \\ \hline       
Input Parameter                 & Value         &  Fit Output    &   Range (${\rm CL} > 5~\%$)                             \\
\hline
$|V_{ud}|$                      & $ 0.97394 \pm 0.00089$                       &  $\lambda$                           & $0.2221     \pm 0.0041$                    \\
$|V_{us}|$                      & $ 0.2200  \pm 0.0025$                        & $A$                                  & $0.763      -   0.905$ \\
$|V_{cd}|$                      & $ 0.224   \pm 0.014$                         & $\bar{\rho}$                         & $0.07       -   0.37$ \\
$|V_{cs}|$                      & $ 0.969   \pm 0.058$                         & $\bar{\eta}$                         & $0.26       -   0.49$ \\
$|V_{ub}|$                      & $(3.49    \pm 0.24  \pm 0.55)\cdot 10^{-3}$  & $J$($10^{-3}$)                       & $ 2.2       -   3.7$ \\
$|V_{cb}|$                      & $(40.4    \pm 1.3   \pm 0.9)\cdot 10^{-3}$   & $\sin{2\alpha}$                      & $-0.90      -   0.51$ \\
$\Delta m_{d}$                  & $(0.489   \pm 0.008)~{\rm ps}^{-1}$          & $\sin{2\beta}$                       & $ 0.59      -   0.88$ \\
$\Delta m_{s}$                  & Amplitude Spectrum                           & $\gamma$                             & $37^{\circ} -   80^{\circ}$ \\
$|\epsilon_{K}|$                & $(2.271   \pm 0.017) \cdot10^{-3}$           & $\Delta m_{\rm s}~({\rm ps}^{-1})$   & $14.6       - 32.0$\\
$\sin{2\beta}$                  & $0.793    \pm 0.102$                         & $f_{B_{d}}\sqrt{B_{d}}~({\rm MeV})$  & $192        - 284$ \\
$m_t(\overline{\rm MS})$	        & $(166     \pm 5)~{\rm GeV}$                  & $B_{K}$                              & $0.52       - 1.68$\\
$m_c~({\rm GeV})$		& $1.30     \pm 0.10$                          & $m_{t}~({\rm GeV})$                  & $95         - 405$ \\
$B_K$                           & $0.87     \pm 0.06  \pm 0.13$                & ${\cal B}(K_{L} \rightarrow \pi^{0}  \nu \bar{\nu})\cdot(10^{-11})$ & $1.6 -  4.4$ \\
$\eta_{cc}$                     & $1.38     \pm 0.53$                          & ${\cal B}(K^{+} \rightarrow \pi^{+}  \nu \bar{\nu})\cdot(10^{-11})$ & $4.8 -  9.4$ \\
$f_{B_d}\sqrt{B_d}~({\rm GeV})$ & $(0.230   \pm 0.028 \pm 0.028)~{\rm GeV}$    & ${\cal B}(B^{+} \rightarrow \tau^{+} \nu_{\tau})\cdot(10^{-7})$     & $5.6 - 23.8$ \\
$\eta_{B}(\overline{\rm MS})$       & $0.55     \pm 0.01$                          & ${\cal B}(B^{+} \rightarrow \mu^{+}  \nu_{\mu})\cdot(10^{-5})$      & $2.2 -  9.4$ \\
$\xi$                           & $1.16     \pm 0.03  \pm 0.05$                & 
                                        &               \\
\hline
\end{tabular}
\caption{Left: input values for the global fit. Right: fit results 
quoted for ${\rm CL}>5~\%$.
\label{tab:measurements}}
\end{table}
If the hypothesis ``the CKM picture of the SM is correct'' is 
accepted, CLs in parameter subspaces $a$, {\it e.g.} 
$a = (\bar{\rho},\bar{\eta})$\,\cite{Wolfenstein}, are evaluated. 
For fixed $a$, one calculates
$\Delta \chi^2(a) 
= \chi^2_{\rm min; \mu}(a) - \chi^2_{{\rm min};y_{\rm mod}}$,
where $\mu$ stands for all model parameters (including 
$y_{\rm QCD}$) with the exception of $a$. The corresponding 
CL is obtained from
${\rm CL}(a) = {\rm Prob}(\Delta \chi^2(a),N_{\rm dof})$,
where $N_{\rm dof}$ is the number of degrees of freedom, 
in general the dimension of the subspace $a$. Since 
the CL depends on the choice of the ranges for the 
$y_{\rm QCD}$, the results obtained in the fit have to 
be interpreted with care.
\begin{figure}
\includegraphics[height=.335\textheight]{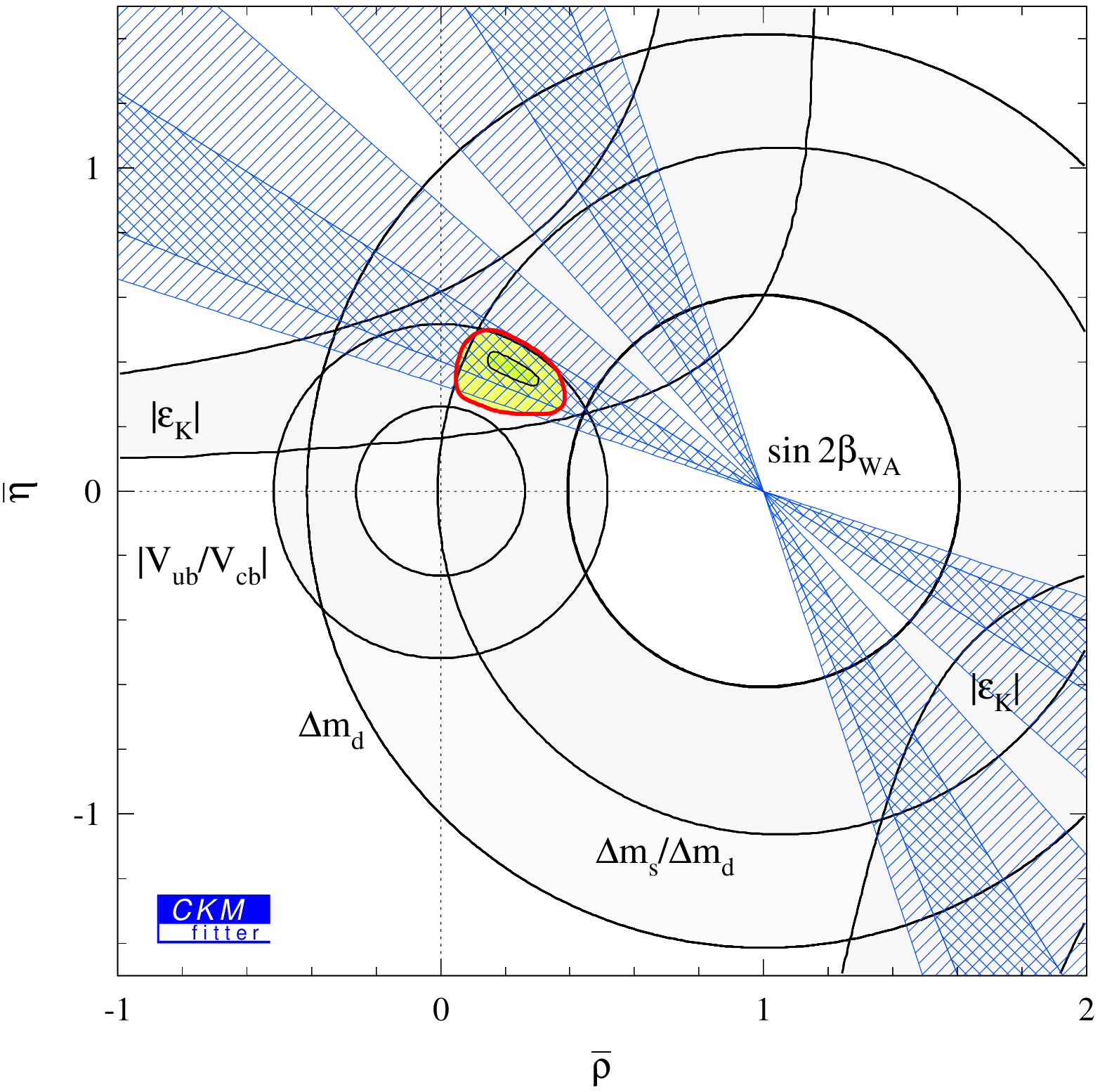}
\includegraphics[height=.335\textheight]{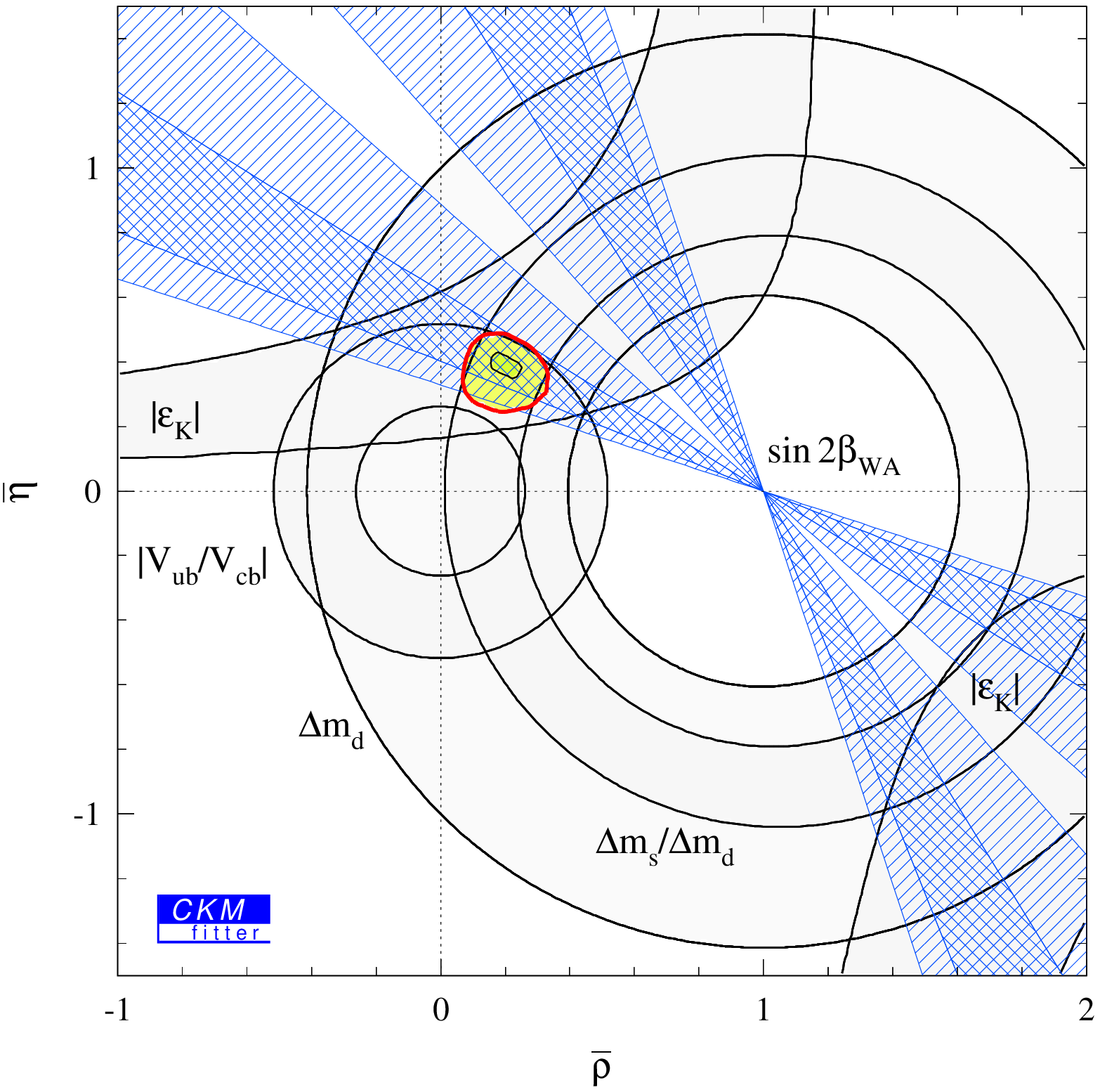}
\caption{Left: constraints in the $(\bar{\rho},\bar{\eta})$ 
plane where $\bar{\rho}$ and $\bar{\eta}$ are defined by 
$\bar{\rho}=\rho(1-\lambda^{2}/2)$, 
$\bar{\eta}=\eta(1-\lambda^{2}/2)$. Shown are the $5~\%$ CLs 
for the individual constraints as shaded areas and the 
$1\sigma$- and $2\sigma$- contours from $\sin{2\beta}$. In 
addition, the $5~\%$ and $90~\%$ CL contours for the combined 
fit are drawn. Right: constraints in the $(\bar{\rho},\bar{\eta})$ 
plane if the likelihood ratio is used for the $\Delta m_{s}$ 
constraint.
\label{rhoeta_withs2b_wa}}
\end{figure}
\\
The input values used in this analysis are listed in 
Tab.~\ref{tab:measurements}. For $|V_{ub}|$, inclusive 
measurements from LEP and exclusive measurements from 
CLEO have been used. The preliminary CLEO lepton endpoint 
analysis\,\cite{CLEOsemileptonic} using moments obtained 
from $B \rightarrow X_s \gamma$ is not yet included. For 
$|V_{cb}|$, inclusive measurements from LEP, the 
measurements of $B \rightarrow D^* \ell \nu$ at zero-recoil 
and the moments analysis from CLEO\,\cite{CLEOsemileptonic}, 
using inclusive $B \rightarrow X_c \ell \nu$ and 
$B \rightarrow X_s \gamma$ decays, have been combined. 
The uncertainty on $\Delta m_{d}$ has been significantly 
reduced due to the measurements from the $B$-factories\,\cite{sin2beta}. 
However, the constraint on ($\bar{\rho},\bar{\eta}$) is not 
improved since it is dominated by the theoretical uncertainty 
on $f_{B_{d}}\sqrt{B_{d}}$.
For $\Delta m_{s}$, the most recent combined amplitude spectrum 
from\,\cite{LEPB} is included in the fit using a modified version 
of the standard amplitude method\,\cite{HLLL}. If the amplitude 
spectrum is translated into a likelihood 
ratio\,\cite{RoussarieMoser,Ciuchini}, a stronger constraint 
is obtained.
However, to our knowledge, it has not been demonstrated so far
that the likelihood ratio can be interpreted as a probability 
density function. Hence, we use the more conservative method of 
Ref.\,\cite{HLLL} for the numerical analysis presented here. 
For $\sin\!{2\beta}$, the world average is used. It 
should be noted that the most precise measurements from 
BABAR and Belle\,\cite{sin2beta} differ presently by 
about two standard deviations.
\section{Fit Results}\label{sec:results}
The global minimum of the CKM fit is found to be
$\chi^2_{{\rm min};y_{\rm mod}}=2.3$, resulting in a 
goodness-of-fit of $71\%$. It quantifies the excellent 
agreement between experimental data and the CKM picture 
of the SM.
Fig.~\ref{rhoeta_withs2b_wa} shows the $(\bar{\rho},\bar{\eta})$ 
plane. Drawn are $5\%$ CL contours from the single constraints 
using $\Delta m_{d}$, $\Delta m_{s}$, $|V_{ub}/V_{cb}|$ and 
$|\epsilon_{K}|$, respectively, and the $1\sigma$- and 
$2\sigma$-contours for the four-fold ambiguity on $\beta$ from 
$\sin{2\beta}$. 
Shown in addition are the contours for the combined fit 
including $\sin\!{2\beta}$. The statistical precision of the 
$\sin\!{2\beta}$ measurement already competes with the indirect, 
theoretically limited constraints. Fig.~\ref{rhoeta_withs2b_wa} 
(right) illustrates the improved constraint when using the 
likelihood ratio for $\Delta m_{s}$. Selected numerical results 
are summarized in Tab.~\ref{tab:measurements}.
\vspace{-0.2cm}
\section{Charmless Two Body Decays}\label{sec:charmless}
\begin{figure}
\includegraphics[height=.335\textheight]{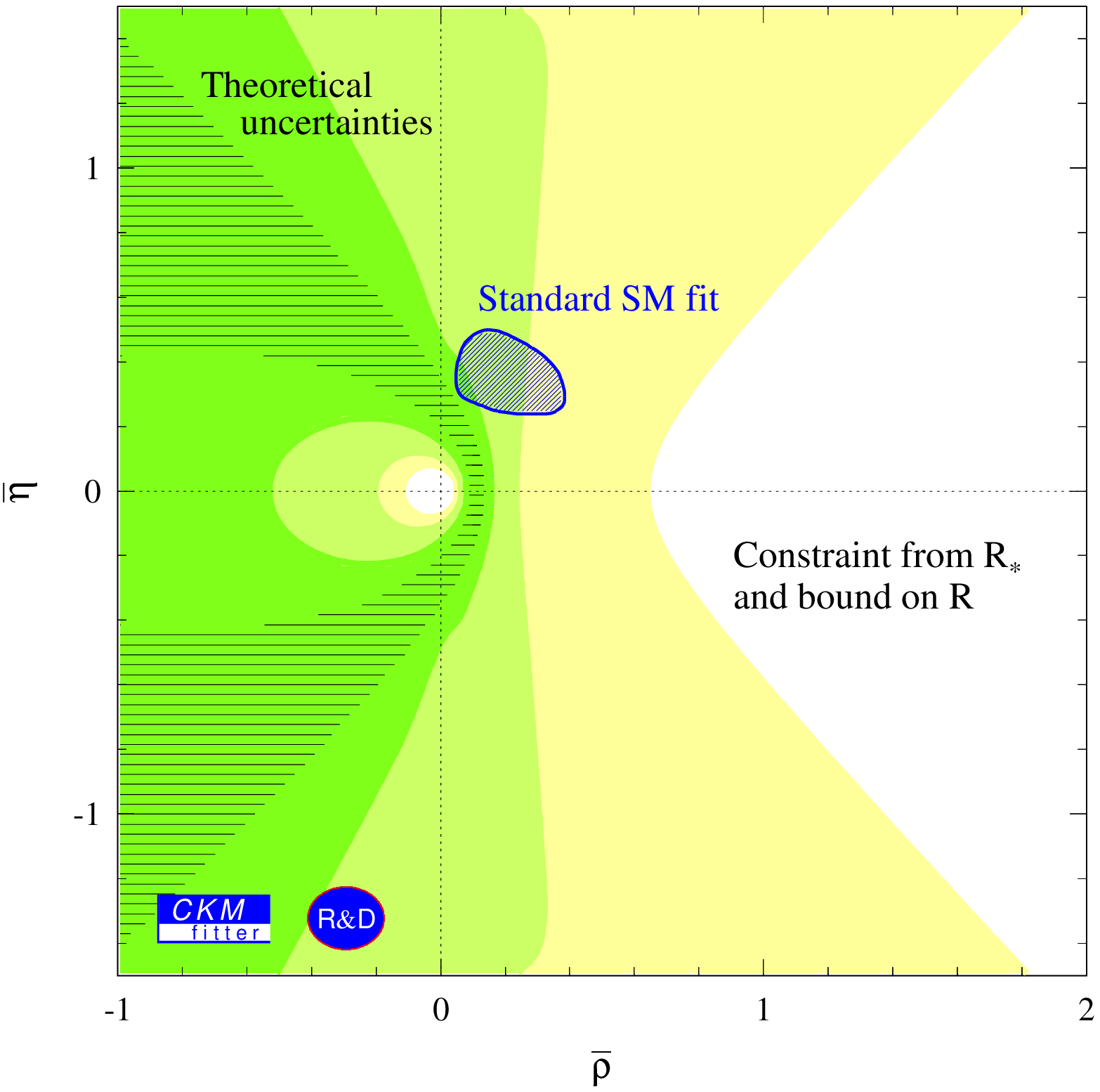}
\includegraphics[height=.335\textheight]{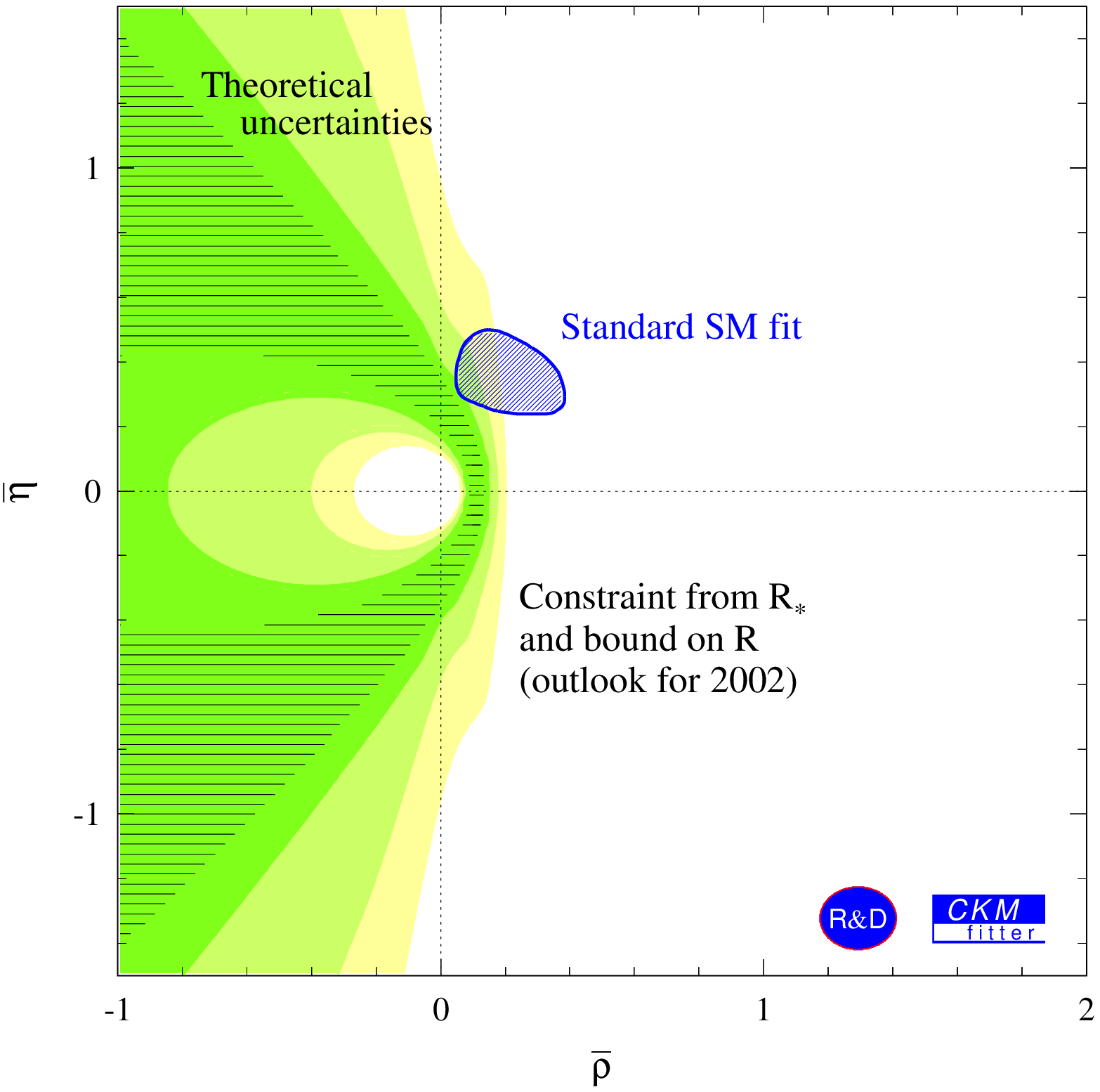}
\caption{Left: Constraints in the $(\bar{\rho},\bar{\eta})$ 
plane using $R_{*}$ (see text) and the bound on $R$. The 
CL is indicated by the shaded areas; dark-grey: $> 90~\%$, 
gray: $90~\%-32~\%$, light-gray: $32~\%-5~\%$. 
Right: Hypothetical constraints for summer 2002. 
\label{nrfm}}
\end{figure}
\begin{figure}
\includegraphics[height=.335\textheight]{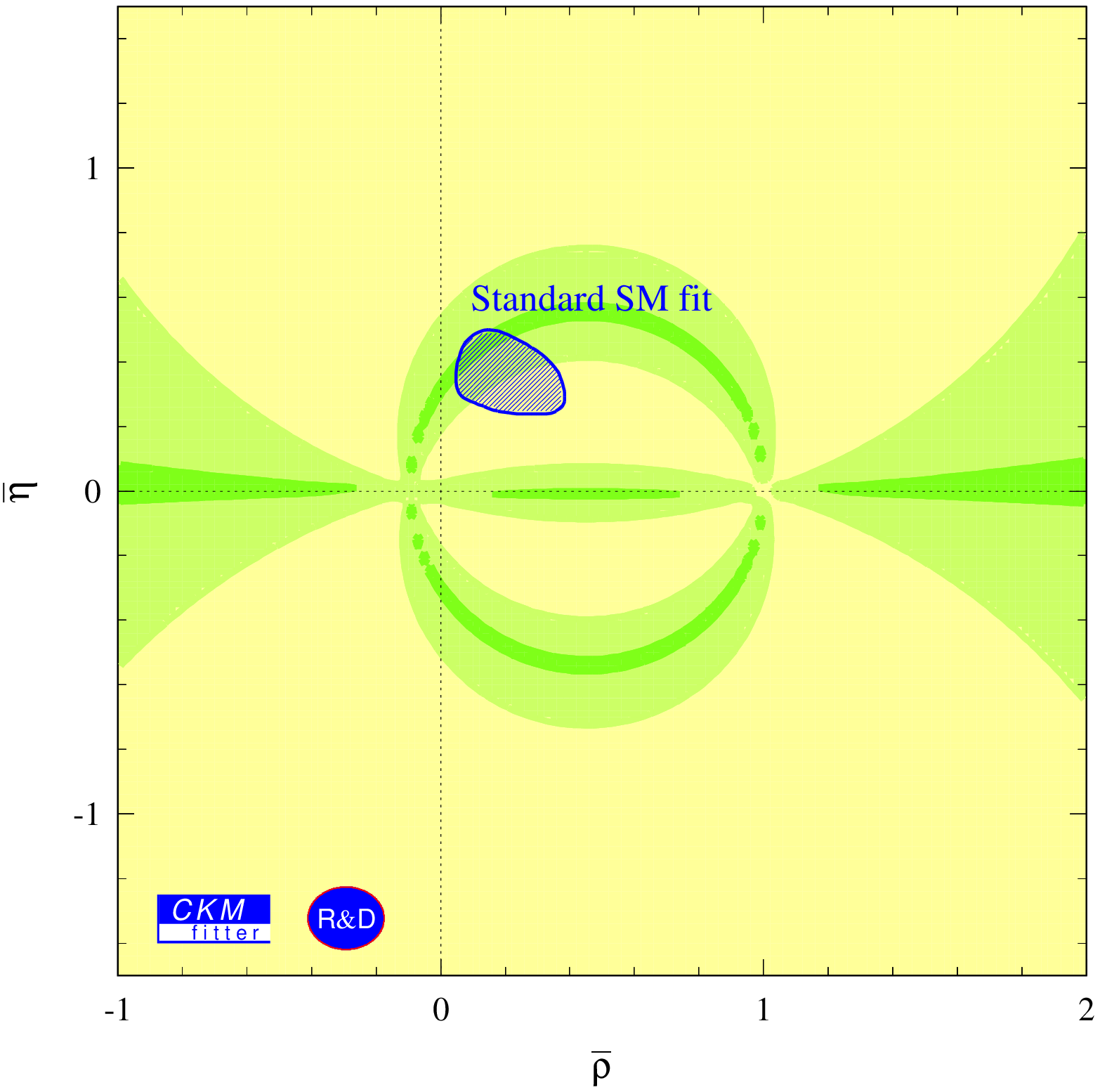}
\includegraphics[height=.335\textheight]{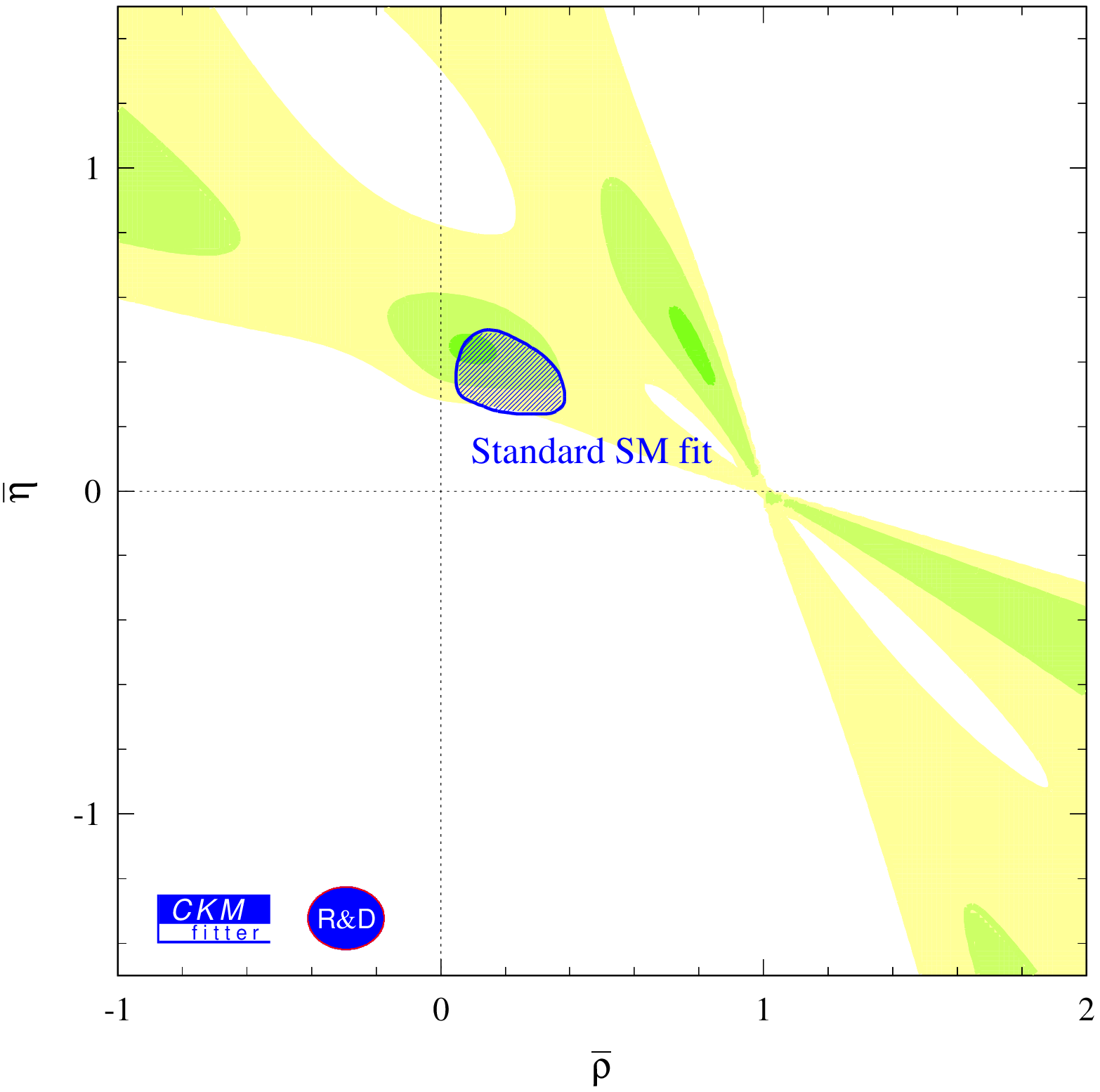}
\caption{Left: Constraints on $(\bar{\rho},\bar{\eta})$ from
$S_{\pi\pi}$ using the penguin-over-tree ratio from FA.
Right: Constraints when using also $\sin\!{2\beta}$.  
\label{s2as2b}}
\end{figure}

Constraints on the angle $\gamma$ can be obtained from 
$B \rightarrow \pi\pi,K\pi$ decays. 
Based on color 
transparency arguments, theoretical calculations such 
as the QCD Factorisation Approach (FA)\,\cite{BBNS} and 
the QCD hard scattering approach\,\cite{KLS} have been 
developed.
Recently, the FA has been implemented in CKMfitter. At 
present, it is premature to infer reliable constraints 
on the basis of these calculations due to open theoretical 
questions\,\cite{BBNS,KLS,CFMPS}. Data from the $B$-factories 
are not yet precise enough to probe the calculations in detail. 
Hence, all fit results are marked by an appropriate 
``${\rm R \& D}$'' logo.
\\
For this review, a global fit to $B \rightarrow \pi\pi,K\pi$ 
branching fractions and direct CP asymmetries measured in 
self-tagging $B \rightarrow K\pi$ decays has been performed 
within the framework of the FA, where most recent experimental
results  from BABAR\,\cite{BABAR}, Belle\,\cite{BELLE} and 
CLEO\,\cite{CLEO} have been used. The numerous theoretical 
parameters are let free to vary within the ranges given in 
Ref.\,\cite{BBNS}. We find $\chi^2_{{\rm min};y_{\rm mod}}=2.0$ 
and conclude that data are consistently described within the FA. 
The best FA fits are found at $\gamma \approx 80^{\circ}$ and 
are in agreement with the constraints from the standard fit.
\\
Using less theoretical assumptions, ratios of branching fractions 
can be formed to derive constraints in the $(\bar{\rho},\bar{\eta})$ 
plane. As an example, the CP-averaged ratio
\begin{eqnarray}
R = \frac{\tau_{B^{\pm}}}{\tau_{B^{0}}}
    \frac{{\cal B}(B^{0} \rightarrow K^{\pm} \pi^{\mp})}
         {{\cal B}(B^{\pm}   \rightarrow K^{0}   \pi^{\pm})}~,
\end{eqnarray}
provides the bound $R > \sin^{2}\gamma$ which is independent of the 
strong phases\,\cite{FM}. Unfortunately, the present world average 
$R = 1.07^{+0.19}_{-0.15}$
leads to weak constraints only owing to the tails of the experimental 
errors.
The ratio
\begin{eqnarray}
R_{*} = \frac{{\cal B}(B^{\pm} \rightarrow K^{0} \pi^{\pm})}
             {2 \cdot {\cal B}(B^{\pm} \rightarrow K^{\pm} \pi^{0})}~,
\end{eqnarray}
measured to be $R_{*} = 0.70^{+0.16}_{-0.13}$, can be 
used to derive bounds in the $(\bar{\rho},\bar{\eta})$ 
plane\,\cite{NR}. An important input for the theoretical prediction 
of $R_{*}$ is the tree-to-penguin ratio (P/T) which can be determined 
experimentally using the relation
\begin{eqnarray}
  \bar{\epsilon}_{3/2} 
= R_{\rm th} \cdot \tan \theta_{\rm C} \cdot \frac{f_{K}}{f_{\pi}}
\sqrt{\frac{2\cdot {\cal B}(B^{\pm} \rightarrow \pi^{\pm} \pi^{0})}
                  {{\cal B}(B^{\pm} \rightarrow   K^{0} \pi^{\pm})}~,
     }
\end{eqnarray}
where $R_{\rm th}$ stands for SU(3) breaking corrections 
estimated in the FA to be\,\cite{BBNS} 
$R_{\rm th} =0.98 \pm 0.05$. The bound on $R_{*}$ can be 
translated into a prediction if additional information on the 
strong phases is inserted\,\cite{BBNS}. Adopting the values for 
theoretical ranges quoted in Ref.\,\cite{BBNS}, one obtains the 
constraints shown in Fig.~\ref{nrfm}. At present, the constraints 
remain rather weak due to the limited experimental precision. 
The slight deformation of the shape pattern around 
$\gamma \approx 90^{\circ}$ is due to the bound on $R$. 
For summer 2002, an integrated luminosity of $100~{\rm fb^{-1}}$ 
is expected to be collected by each experiment, BABAR and Belle. 
The experimental precision will then start to provide interesting 
constraints, as can be seen from Fig.~\ref{nrfm} obtained 
assuming the present central experimental values and appropriately
rescaling their errors. However, the constraints would still rely 
on the validity of some theoretical assumptions not yet fully 
explored.
\\
Within the FA, the P/T ratio for $B \rightarrow \pi^{+}\pi^{-}$ 
is predicted. Compared to the present experimental error on the 
time-dependent asymmetry $S_{\pi\pi}=0.03^{+0.53}_{-0.56}\pm0.11$ 
from BABAR, the quoted theoretical uncertainty is much 
smaller\,\cite{BBNS}. In Fig.\,\ref{s2as2b} (left), the constraints 
in $(\bar{\rho},\bar{\eta})$ from $S_{\pi\pi}$ are shown using P/T 
from FA where theoretical uncertainties have been neglected. The 
right plot shows the constraints when also using $\sin\!{2\beta}$.  

\begin{theacknowledgments}
It was a pleasure to attend this conference which was very 
successful despite the tragic events of the 11th September. 
We are indebted to Martin Beneke and Matthias Neubert for 
their help implementing the QCD Factorisation Approach in 
CKMfitter. HL was supported by the Fifth Framework Programme 
of the European Community Research under the grant No. 
HPMF-CT-1999-00032.
\end{theacknowledgments}

\end{document}